\documentclass[aps,twocolumn,prl,showpacs,preprintnumbers,superscriptaddress,amsmath,amssymb,floatfix]{revtex4}
\usepackage{graphicx}% Include figure files
\usepackage{dcolumn}% Align table columns on decimal point
\usepackage{bm}% bold math
\usepackage[english]{babel}

\begin{document}
\preprint{APS/123-QED}

\author{J. G. Donath}
\affiliation{Max-Planck-Institute for Chemical Physics of Solids,
D-01187 Dresden, Germany}
\author{P. Gegenwart}
\affiliation{I. Physik. Institut, Georg-August-Universit\"{a}t
G\"ottingen, D-37077 G\"ottingen}
\author{F. Steglich}
\affiliation{Max-Planck-Institute for Chemical Physics of Solids,
D-01187 Dresden, Germany}
\author{E. D. Bauer}
\author{J. L. Sarrao}
\affiliation{Los Alamos National Laboratory, Los Alamos, NM 87545, USA}

\title{Dimensional crossover of quantum critical behavior in CeCoIn$_5$}
\date{%TCIMACRO{\TeXButton{Date}{\today} }
%BeginExpansion
\today%
%EndExpansion
}

\begin{abstract}
The nature of quantum criticality in CeCoIn$_5$ is studied by
low-temperature thermal expansion $\alpha(T)$. At the field-induced
quantum critical point at $H=5$~T a crossover scale $T^\star\approx
0.3$~K is observed, separating $\alpha(T)/T\propto T^{-1}$ from a
weaker $T^{-1/2}$ divergence. We ascribe this change to a crossover
in the dimensionality of the critical fluctuations which may be
coupled to a change from unconventional to conventional quantum
criticality. Disorder, whose effect on quantum criticality is
studied in CeCoIn$_{5-x}$Sn$_x$ ($0\leq x\leq 0.18$), shifts
$T^\star$ towards
higher temperatures.% and stabilizes conventional quantum criticality.
\end{abstract}

\pacs{71.10.Hf,71.27.+a,74.70.Tx} \maketitle

Quantum criticality in heavy fermion (HF) systems continues to
attract interest due to the occurrence of highly anomalous metallic
states with severe deviations from Landau Fermi liquid (LFL)
behavior \cite{Stewart,Nat Phys} and the emergence of unconventional
superconductivity in close vicinity to antiferromagnetic (AF)
quantum critical points (QCPs) \cite{Mathur}. Neither the nature of
the non-Fermi liquid (NFL) normal state related to quantum
criticality, nor the superconducting (SC) pairing mechanism has been
clarified up to now. It is thus of great interest to investigate
whether quantum criticality in these systems can be described by
conventional theory within the framework of a spin-density-wave
(SDW) instability \cite{Millis,Moriya}, or whether unconventional
scenarios in which the f-electrons localize at the magnetic QCP due
to a destruction of the Kondo resonance \cite{Si,Coleman,Senthil}
may be more appropriate. For the formation of the latter, magnetic
frustration leading to a reduced dimensionality of the critical
fluctuations may be crucial.

%Particular important is the question under which conditions the
%thermodynamic behavior may be in accordance with the predictions of
%the itinerant spin-density-wave theory of quantum criticality
%\cite{Hertz,Moriya,Millis} or, when the latter fails, more drastic
%deviations from LFL behavior occur. For the latter case, a
%decomposition of heavy quasiparticles due to the disappearance of
%the Kondo resonance at the QCP is discussed \cite{Coleman,Si}. This
%may lead to a localization of the f-electrons at the QCP.

The CeMIn$_5$ (M=Rh, Ir, Co) systems are prototypical as they
display a generic phase diagram with unconventional HF
superconductivity in close vicinity to an AF QCP \cite{review}. They
crystallize in a tetragonal structure which can be viewed as an
alternating series of CeIn$_3$ and MIn$_2$ layers. As a result of
the layered crystal structure, the Fermi surface displays a strongly
two-dimensional (2D) character with cylindrical sheets along the
crystallographic $c$-axis \cite{dhva}. Compared to cubic CeIn$_3$, a
HF superconductor with $T_c=0.2$~K \cite{Mathur} in a very narrow
pressure range close to the magnetic quantum phase transition at
$p_c\approx 2.6$\,GPa, SC transition temperatures of about 2~K are
observed over wide pressure ranges for the tetragonal CeRhIn$_5$ (at
$p\geq 1.6$~GPa) and CeCoIn$_5$ (at ambient pressure)
\cite{Hegger,Petrovic}. This $T_c$ enhancement has been attributed
to the layered crystal structure and, relatedly, strongly
anisotropic magnetic fluctuations \cite{Petrovic}. Indeed the
nuclear magnetic relaxation rate $1/T_1$ of CeCoIn$_5$ displays a
weak $T^{1/4}$ dependence in the normal state between 2 and 40~K
which signals strongly anisotropic quantum critical fluctuations
\cite{Kawasaki}.

\begin{figure}
\centerline{\includegraphics[width=\linewidth,keepaspectratio]{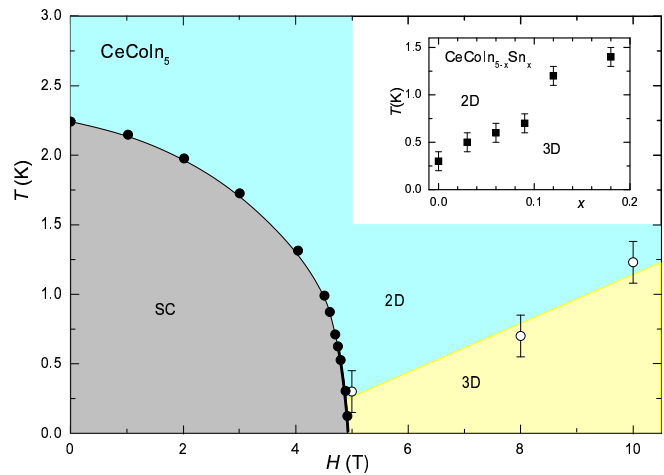}}
\caption{(Color online) Phase diagram of CeCoIn$_5$ for $H\parallel
c$ as determined from thermal expansion. Superconducting phase in
gray with first-order boundary below 0.7~K indicated by thick black
line. Regions where thermal expansion follows 2D- and 3D quantum
critical behavior are marked in blue and yellow, respectively.
%Solid diamonds indicate crossover found in previous specific heat
%experiments \cite{Bianchi nfl}.
The inset displays the evolution of the crossover with Sn-doping in
CeCoIn$_{5-x}$Sn$_x$ at the respective $H_{c2}(x)$.} \label{fig1}
\end{figure}

The aim of this Letter is a detailed investigation of the nature of
quantum criticality in CeCoIn$_5$. We focus in particular to the
region very close to the upper critical field $H_{c2}$ for
superconductivity (5~T for $H\parallel c$, cf. Figure 1) which
previously has been studied by heat and charge transport
\cite{Paglione 2003,Paglione 2006,Tanatar 2007} and specific heat
measurements \cite{Bianchi nfl}. Diverging coefficients of the $T^2$
contributions to the electrical and the thermal resistivity prove
the existence of a magnetic-field induced QCP at 5~T. NFL behavior
in the temperature dependence of the electronic specific heat
coefficient at 5~T has been described in the frame of the SDW theory
\cite{Bianchi nfl}. However, below 0.3~K a large nuclear
contribution arising from the Zeeman splitting of In-nuclear moments
needs to be subtracted. Therefore, the data do not allow to
distinguish between a saturation or logarithmic divergence at lowest
temperatures and thus further {\it thermodynamic} measurements are
needed to determine the nature of quantum criticality in the system
(transport data will be discussed later).

%Upon cooling to temperatures below 1~K, the electronic specific heat
%coefficient $\Delta C(T)/T$ increases logarithmically. Within the
%SDW theory for an AF QCP, such a temperature dependence would imply
%the existence of 2D quantum critical fluctuations
%\cite{Millis,Moriya}. At temperatures below 0.1~K, the larger
%scatter of the data, caused by the subtraction of a dominating
%nuclear specific heat contribution, does not allow to distinguish
%between (i) a logarithmic divergence (cf. dotted line in Figure 1a),
%(ii) a square-root dependence, expected for a 3D AF QCP (dashed
%line) \cite{Millis,Moriya}, (iii) or saturation indicating LFL
%behavior. Thus, further thermodynamic measurements are needed to
%determine the nature of quantum criticality in the system.

Thermal expansion is ideally suited for this purpose. It probes the
pressure dependence of the entropy which close to QCPs is accumulated at
finite temperatures. Scaling arguments have revealed that thermal
expansion $\alpha(T)$ is far more singular than specific heat $C(T)$ in
the approach of any pressure-sensitive QCP \cite{zhu}. Within the SDW
theory the leading contribution to $\alpha(T)/T$ diverges like
$T^{-1/2}$ and $T^{-1}$ for 3D and 2D AF QCPs, respectively \cite{zhu}.
Both can easily be distinguished from $\alpha(T)/T=const.$ expected for
a LFL. Especially important in this context, thermal expansion, in
contrast to specific heat, is not affected by nuclear hyperfine
contributions.

%\begin{figure}
%\centerline{\includegraphics[width=\linewidth,keepaspectratio]{Figure1.eps}}
%\caption{(Color online) a: Electronic specific heat coefficient
%$\Delta C/T$ of CeCoIn$_5$ in a field of 5~T applied parallel to the
%$c$-axis vs $T$ (on logarithmic scale) \cite{Bianchi nfl}. Red solid
%line indicates a fit to Moriya's spinfluctuation theory
%\cite{Bianchi nfl}. Dotted and dashed lines represent $\log T$ and
%$const-\sqrt{T}$ dependences, respectively. b: Corresponding linear
%thermal expansion coefficient along the $c$-axis as $\alpha/T$ vs
%$\log T$. Solid line displays $1/T$ dependence. Arrow indicates
%lower limit for $1/T$ dependence.} \label{fig1}
%\end{figure}

For our study, we have used high-quality single crystals of
CeCoIn$_{5-x}$Sn$_x$ grown from In flux, whose low-temperature
specific heat and electrical resistivity are discussed in
\cite{Bianchi nfl,Bauer 2005,Bauer 2006}. For details on the sample
characterization see \cite{Bauer 2005,Bauer 2006,Daniel 2005}.
Thermal expansion has been determined with the aid of
high-resolution dilatometers at temperatures down to 0.04~K and in
magnetic fields up to 10~T. We have measured the length change
$\Delta L_c$ along the $c$-axis and determined the linear ($c$-axis)
expansion coefficient $\alpha=\partial \ln L_c/\partial T$.

\begin{figure}
\centerline{\includegraphics[width=\linewidth,keepaspectratio]{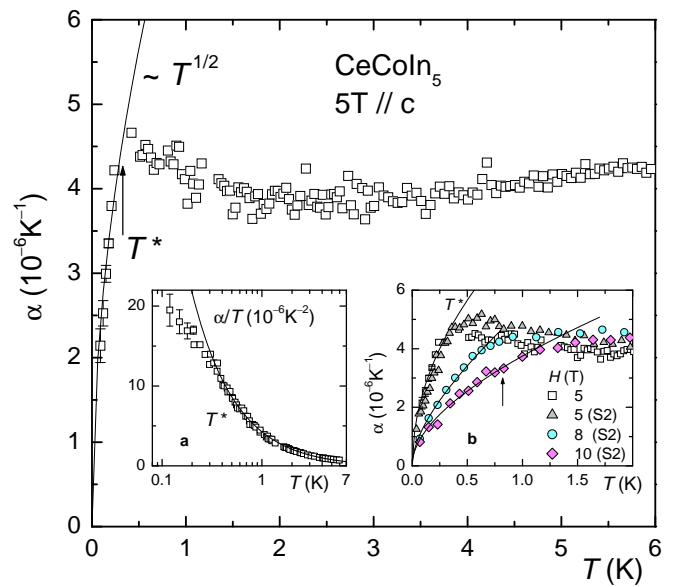}}
\caption{(Color online) Temperature dependence of the linear thermal
expansion coefficient of CeCoIn$_5$ at $H=5$~T ($\parallel c$).
Dotted line and arrow indicate $\alpha\propto \sqrt{T}$ and
crossover temperature $T^\star$, defined as upper limit for this
$T$-dependence, respectively. Inset (a) displays data from main part
as $\alpha/T$ vs $T$ (on logarithmic scale). Solid line indicates
$T^{-1}$ dependence. Inset (b) compares data from main part in the
low-temperature regime with 5, 8 and 10T data obtained from a second
sample (S2). Lines display square-root behavior.} \label{fig2}
\end{figure}

Figure 2 displays our thermal expansion data on undoped CeCoIn$_5$.
At the upper critical field of 5~T, the thermal expansion
coefficient $\alpha(T)/T$ (cf. inset a) grows much stronger upon
cooling than the respective specific heat coefficient which diverges
only logarithmically \cite{Bianchi nfl}. Over more than one decade
in temperature, i.e. for $0.3{\rm ~K}\leq T\leq 6 {\rm ~K}$, the
data follow a $1/T$ divergence which hints at 2D AF quantum critical
fluctuations \cite{zhu}. The latter may result from the layered
crystal structure \cite{Petrovic,Kawasaki}. At $T^\star\approx
0.3$~K, the temperature dependence changes to $\alpha\propto
T^{1/2}$ (see main part and inset b, which also displays data
obtained on a second sample down to 40~mK). Note, that $\alpha/T$
does not show a saturation excluding the formation of a LFL above
the lowest measured temperature. The square-root behavior for
$\alpha(T)$ is compatible with a 3D AF QCP of itinerant nature
\cite{zhu} as observed for CeNi$_2$Ge$_2$ \cite{Nat Phys} and
CeIn$_{3-x}$Sn$_x$ \cite{Kuechler 2006}.

As $H$ is increased above 5~T, $T^\star$ increases and $\alpha(T)$
becomes less singular, i.e. the coefficient of the square-root
contribution decreases (cf. Fig.2, inset b). This suggests that the
system is tuned away from the QCP, compatible with previous studies
\cite{Paglione 2003,Bianchi nfl}, although LFL behavior is not yet
fully established in thermal expansion.
%Note, that specific-heat measurements have revealed a broad
%crossover regime in the $T-H$ phase diagram between NFL-behavior
%$C_{el}/T\propto \log T$ and LFL behavior $C_{el}/T \approx const$
%\cite{Bianchi nfl}. However, because specific heat is much less
%singular than thermal expansion, a 2D to 3D crossover could not be
%resolved in the former.

\begin{table}
  \centering
  \begin{tabular}{cccc}\\\hline\hline
    $x$ & $T_c$ (K) & $H_{c2}$ (T) \\\hline
 $0.00$ & $(2.25\pm 0.05)$ & $(4.9\pm 0.1)$ \\
 $0.03$ & $(1.80\pm 0.05)$ & $(4.5\pm 0.1)$ \\
 $0.06$ & $(1.50\pm 0.05)$ & $(3.9\pm 0.1)$ \\
 $0.09$ & $(1.15\pm 0.05)$ & $(3.4\pm 0.1)$ \\
 $0.12$ & $(0.75\pm 0.05)$ & $(2.5\pm 0.1)$ \\
 $0.18$ & $0$ & $0$ & \\\hline\hline
  \end{tabular}
  \caption{Values for the SC transition temperature $T_c$ and upper critical magnetic field $H_{c2}$ for CeCoIn$_{5-x}$Sn$_x$
  \cite{Bauer 2005}.}
  \label{tab:1}
\end{table}

Our data on CeCoIn$_5$ are summarized in the main part of Fig.~1. We
have observed a crossover scale separating 2D from 3D quantum
critical behavior. To provide further evidence for this crossover
and to investigate how it is influenced by weak disorder, we now
focus on the series CeCoIn$_{5-x}$Sn$_x$ where the Sn-atoms
preferentially occupy the In-1 position within in the tetragonal
plane \cite{Daniel 2005}. Sn doping weakens superconductivity,
leading to a linear suppression of $T_c$ towards zero for $x=0.18$
\cite{Bauer 2005}. The temperature-magnetic field phase diagram of
various CeCoIn$_{5-x}$Sn$_x$ single crystals has previously been
studied by low-temperature electrical resistivity and specific heat
measurements \cite{Bauer 2005,Bauer 2006}. As $T_c$ is reduced, a
corresponding reduction of $H_{c2}$ is observed (for the
$x$-dependence of $T_c$ and $H_{c2}$, see Table I). For all
different Sn concentrations the temperature dependence of the
specific heat displays NFL behavior at the respective upper critical
field and the formation of a LFL state at fields exceeding
$H_{c2}(x)$ \cite{Bauer 2005}. This suggests that field-induced
quantum criticality is always pinned at the upper critical field
$H_{c2}$ when the latter is reduced by Sn doping. Furthermore, the
low-$T$ specific heat coefficient at $H=H_{c2}(x)$ remains unchanged
within the scatter of the data for $0\leq x\leq 0.12$ \cite{Bauer
2005}. On the other hand, the residual resistivity $\rho_0$ shows a
tenfold increase for $x$ ranging from 0 to 0.18, indicating the
effect of disorder scattering due to the random distribution of
Sn-atoms on the in-plane In site \cite{Bauer 2006}. The study of
CeCoIn$_{5-x}$Sn$_x$ thus allows to systematically investigate the
disorder dependence of NFL behavior without tuning the system away
from the QCP.

\begin{figure}
\centerline{\includegraphics[width=\linewidth,keepaspectratio]{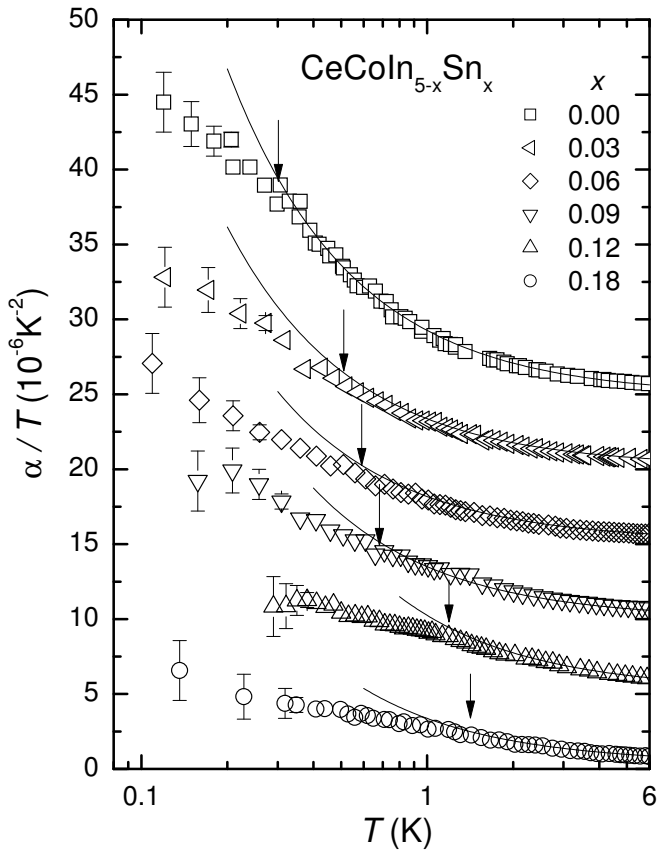}}
\caption{Linear thermal expansion coefficient as $\alpha$ vs. $T$ on a
  logarithmic scale for CeCoIn$_{5-x}$In$_x$ at $H\simeq H_{c2}(x)$
  (given in Table I). Note that data sets are shifted by $5\times
  10^{-6}$~K$^{-2}$, subsequently. Arrows indicate lower limit of
  $\alpha/T\propto T^{-1}$ behavior.} \label{fig3}
\end{figure}

Figure 3 shows $c$-axis thermal expansion data for the various
studied CeCoIn$_{5-x}$Sn$_x$ single crystals at their respective
upper critical magnetic fields (for the zero-field data see
\cite{Donath}). In all these samples, 2D-like quantum critical
behavior $\alpha(T)/T\propto T^{-1}$ is found from 6~K down to a
lower bound which increases from about 0.3~K for $x=0$ to about
1.4~K for $x=0.18$.

\begin{figure}
\centerline{\includegraphics[width=\linewidth,keepaspectratio]{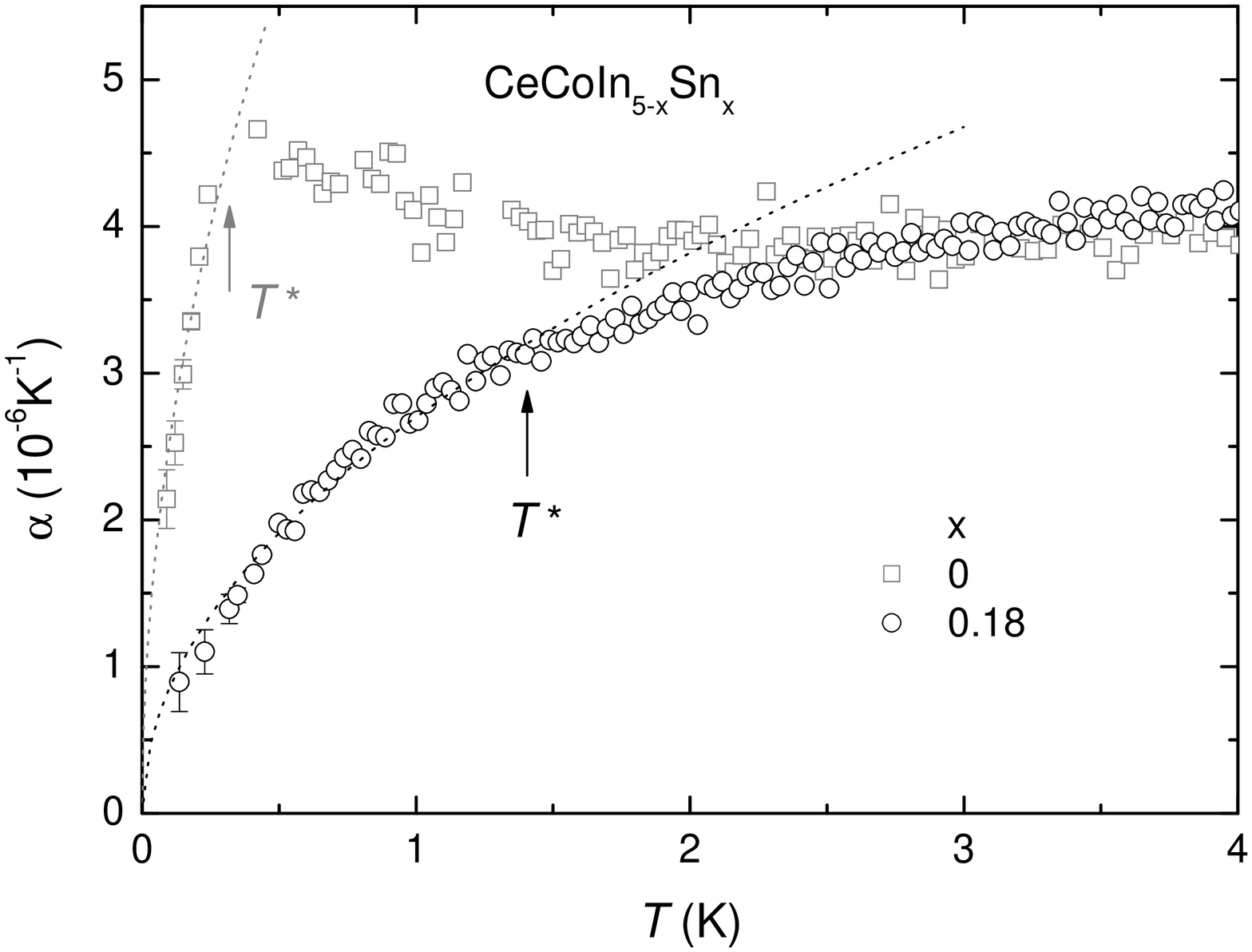}}
\caption{Linear thermal expansion coefficient $\alpha(T)$ of CeCoIn$_5$
  at $H=5$~T ($\parallel c$, open squares), as well as
  CeCoIn$_{4.82}$Sn$_{0.18}$ at $H=0$ (open circles). Dotted lines
  and arrows indicate $\alpha\propto \sqrt{T}$ and crossover
  temperatures $T^\star$, respectively.} \label{fig4}
\end{figure}

Like for $x=0$, the low-$T$ thermal expansion of all samples studied
is well described by $\alpha(T)\propto \sqrt{T}$. For $x=0.18$, this
temperature dependence holds up to $T^\star\approx 1.4$~K, i.e. over
more than one decade (see Fig.~4), providing clear evidence for 3D
AF quantum critical fluctuations in the latter system. Clearly, the
temperature at which the dimensional crossover occurs is shifted
with Sn doping in CeCoIn$_{5-x}$Sn$_x$ to values above 1~K, cf. the
inset of Fig.~1. As stated above, the partial substitution of the
In-(1) site by Sn-atoms enhances impurity scattering without tuning
the system away from the QCP. Our observation of a shift of the
crossover scale $T^\star$ with $x$ is then naturally attributed to
the effect of isotropic impurity scattering, which "smears out" the
magnetic anisotropy. Crossovers have also been observed in the
electrical and heat transport \cite{Paglione 2006,Tanatar 2007} as
well as in the Hall coefficient \cite{Singh 2007} for the current
direction $j\perp c$. However, transport experiments are influenced
by electronic relaxational properties, which can give rise to
complicated behavior for anisotropic and multiband systems like
CeCoIn$_5$. Indeed, for $j\parallel c$ no crossover is visible in
$\rho(T)$, and the Wiedemann-Franz law, which is obeyed for $j\perp
c$, seems to be violated \cite{Tanatar 2007}.

In order to clearly show that, at the QCP in CeCoIn$_{5-x}$Sn$_x$, a
finite energy scale $k_BT^\star$ exists which marks the crossover
from 2D to 3D quantum critical behavior,
%prove the existence of an {\it intrinsic} energy scale near the QCP,
measurements either of the fluctuation spectrum in equilibrium, for
example by inelastic neutron scattering (INS), or of thermodynamic
properties are required. ${\bf q}$-scans of the INS over wide
regions in reciprocal space, required to decide on the
dimensionality of the quantum critical fluctuations, are not
possible at high fields. Therefore, our thermodynamic measurements
provide the only way to investigate this question and indeed prove
such a crossover.

\begin{figure}
\centerline{\includegraphics[width=\linewidth,keepaspectratio]{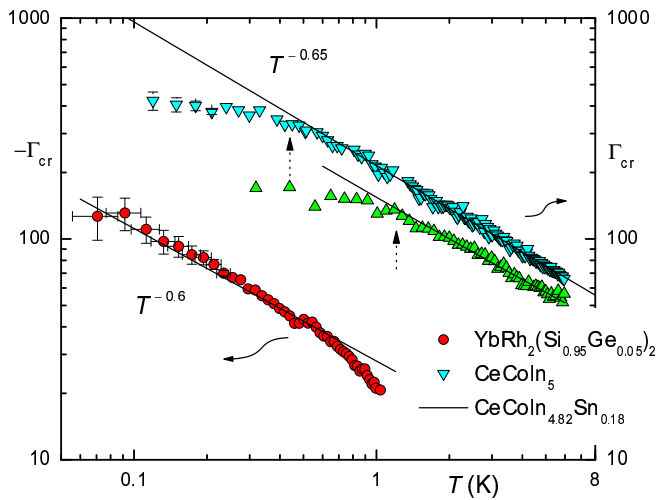}}
\caption{(Color online) Critical Gr\"{u}neisen ratio
$\Gamma_{cr}=V_m/\kappa_T\times \alpha_{cr}/C_{cr}$, where
$\alpha_{cr}$ and $C_{cr}$ denote thermal expansion and specific
heat after subtraction of non-critical background contributions
\cite{zhu}, of YbRh$_2$(Si$_{0.95}$Ge$_{0.05}$)$_2$ (left axis,
\cite{Nat Phys}) and CeCoIn$_{5-x}$In$_x$ ($x=0$ at $H=5$~T and
$x=0.18$ at 0~T, right axis). For the latter, the molar volume and
isothermal compressibility equal $V_m=9.57\cdot 10^{-5}$ m$^3$/mol
and $\kappa_T=(3.43\pm 0.16)\times 10^{-3}$~GPa$^{-1}$ \cite{Kumar},
respectively. A small background term [$0.35\times 10^{-6} {\rm
~K}^{-2}$] has been subtracted from $\alpha/T$ for $x=0.12$. No
specific-heat background contributions have been subtracted since
$C(T)/T\propto \log T$ at $T>T^\star$ ($T^\star$ indicated by dotted
arrows). The so-derived critical Gr\"{u}neisen ratio is invalid for
$T<T^\star$, where the specific heat is dominated by a non-critical
contribution \cite{footnote}. Lines indicate power-law behavior at
$T>T^\star$.} \label{fig5}
\end{figure}

We now address the nature of quantum criticality (SDW-type or
unconventional) in the regime where 2D-like behavior is observed.
Theory suggests that 2D fluctuations are necessary for the
occurrence of locally-critical quantum criticality \cite{Si}. The
latter is well established for the magnetic-field tuned AF QCP in
the heavy fermion system YbRh$_2$Si$_2$ and its slightly Ge-doped
variant YbRh$_2$(Si$_{0.95}$Ge$_{0.05}$)$_2$, for which the critical
field is almost zero \cite{Nat Phys}. It is therefore very
interesting to compare the low-$T$ thermodynamics of
CeCoIn$_{5-x}$Sn$_x$ with the latter system. Of particular
importance is the temperature dependence of the critical Gr\"uneisen
ratio $\Gamma_{cr}$, i.e. the ratio of the critical components of
thermal expansion to specific heat. It has previously been shown,
that $\Gamma_{cr}(T)\propto T^{-\epsilon}$ with $\epsilon=1$ and
$2/3$ for conventional and unconventional quantum criticality,
respectively \cite{Nat Phys}.
%Finally, we address the behavior {\it above} $T^\star$ and compare
%the thermal expansion and Gr\"uneisen ratio of undoped CeCoIn$_5$ at
%the field-tuned QCP near $H=5$~T with corresponding results on
%YbRh$_2$(Si$_{0.95}$Ge$_{0.05}$)$_2$ which displays an AF QCP very
%close to $H\approx 0$ \cite{Kuechler}. For the latter system, a
%stronger than logarithmic mass divergence and fractional exponent in
%the Gr\"uneisen ratio have been observed, both incompatible with the
%itinerant SDW scenario and suggestive of unconventional quantum
%criticality \cite{Kuechler,Custers}.

Figure 5 shows striking similarities in the temperature dependence
of the critical Gr\"uneisen ratio of
YbRh$_2$(Si$_{0.95}$Ge$_{0.05}$)$_2$ and CeCoIn$_{5-x}$Sn$_x$ at
temperatures above $T^\star(x)$, i.e. in the 2D regime:
%Figure 5 compares the low-temperature thermal expansion of this
%latter system with corresponding results on CeCoIn$_5$. Most
%interestingly, above $T^\star$  CeCoIn$_5$ shows a similar $1/T$
%contribution to the expansion coefficient as found in the former
%system. This similarity is also reflected in the analysis of the
%Gr\"uneisen ratio shown in the inset which also includes the
%$x=0.12$ sample.
%Here, for the Yb-system and for $x=0.12$ a normal component, related
%to the constant background term in $\alpha(T)/T$ has been subtracted
%[$0.35\times 10^{-6}$~K$^{-2}$ for the latter]; for undoped
%CeCoIn$_5$, such noncritical contribution is tiny and we have
%analyzed the "raw" data being proportional to the ratio of the total
%thermal expansion to specific heat. Note, that since the specific
%heat coefficient grows at least logarithmically in this temperature
%regime \cite{Bianchi nfl}, no background contributions to the
%specific heat have to be subtracted \cite{zhu,footnote}.
A rather similar fractional Gr\"uneisen exponent is found which is
close to the prediction for the locally-critical QCP scenario in the
presence of xy anisotropy \cite{Si}. Theoretically, it has been
shown that such behavior requires quasi-2D quantum critical
fluctuations \cite{Si} supporting further the latter at $T>T^\star$
in CeCoIn$_5$. In view of the lack of superconductivity in
YbRh$_2$Si$_2$ (at least for $T>10$~mK), it is highly desirable to
check, whether or not a similar crossover towards conventional
behavior at temperatures below the lower limit of previous studies
(20~mK) takes place in the latter material.

%On the other hand, the fact that the WF law is obeyed below
%$T^\star$ proves conventional quantum criticality in the
%low-temperature limit.

%Since for $T<T^\star$ the critical contributions \cite{zhu} to
%specific heat and thermal expansion could be described by $\sqrt{T}$
%(cf. Fig. 1a) and $1/\sqrt{T}$ (cf. Fig. 2), the critical
%Gr\"uneisen ratio in this temperature regime shows indeed a $1/T$
%divergence, compatible with the SDW theory.

To summarize, we have found thermodynamic evidence for a finite
crossover scale $T^\star$ at the magnetic-field tuned QCP in
CeCoIn$_5$. We associate $T^\star$ with a dimensional crossover from
2D ($T>T^\star$) to 3D ($T<T^\star$) quantum critical behavior. The
introduction of disorder shifts the crossover scale towards higher
temperatures.

Stimulating discussions with M. Nicklas, Q. Si and S. Wirth are
gratefully acknowledged. Work at Dresden and G\"{o}ttingen was
partially financed by the DFG Research unit 960 (Quantum phase
transitions), while work at Los Alamos was carried out under the
auspices of the U.S. DOE.


\begin{references}
\bibitem{footnote} For $T<T^\star$, specific heat follows the predictions of the 3D SDW scenario \cite{Bianchi
nfl}, i.e. $C(T)/T=\gamma'-\beta' \sqrt{T}$. Since
$\alpha_{cr}\propto \sqrt{T}$, $\Gamma_{cr}\propto T^{-1}$ (as
expected \cite{zhu}).
\end{references}
\end{document}